\documentstyle[multicol,aps,epsf]{revtex}
\begin{document}
\draft
\title{Phase-Coherent Transport through a Mesoscopic System: A New Probe of Non-Fermi-Liquid Behavior}
\author{Michael R. Geller}
\address{Department of Physics and Astronomy, University of Georgia, Athens, Georgia 30602-2451}
\date{April 21, 1998}
\maketitle

\begin{abstract}
A novel chiral interferometer is proposed that allows for a direct measurement of the phase of 
the transmission coefficient for transport through a variety of mesoscopic structures in a strong 
magnetic field. The effects of electron-electron interaction on this phase is investigated with 
the use of finite-size bosonization techniques combined with perturbation theory resummation. New 
non-Fermi-liquid phenomena are predicted in the fractional quantum Hall effect regime that may be used 
to distinguish experimentally between Luttinger and Fermi liquids.
\end{abstract}

\begin{multicols}{2}

\begin{figure}
\begin{center}
\leavevmode
\epsfbox{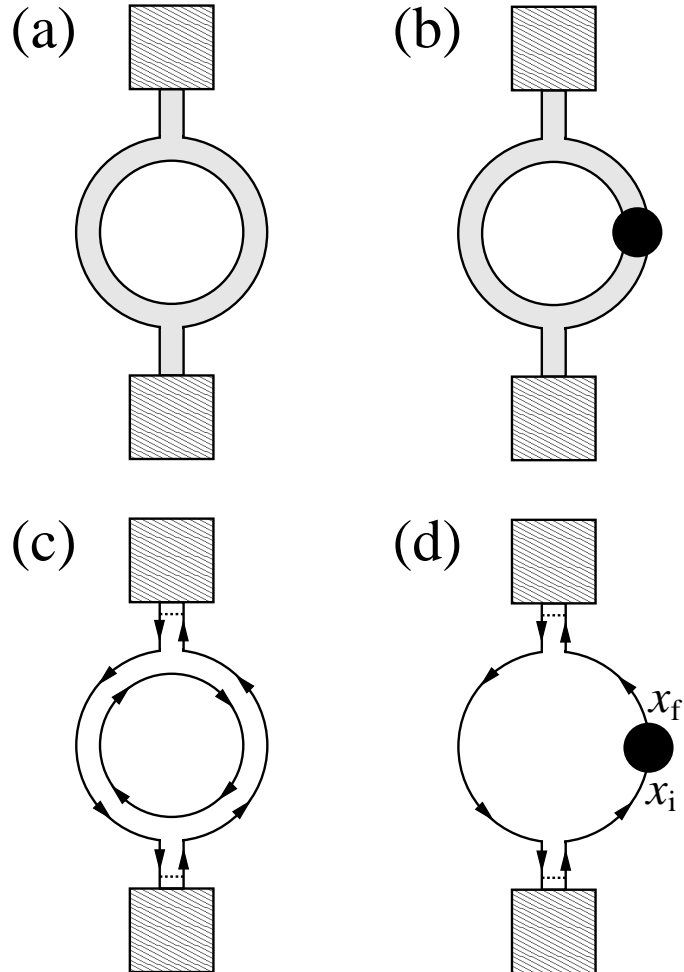}
\caption{(a) Semiconductor interferometer in zero field. A two-dimensional electron gas 
(shaded region) is connected to source and drain contacts. (b) Phase-coherent scatterer (solid black 
circle) with transmission coefficient $t$ inserted into one arm. (c) Interferometer in the quantum Hall 
effect regime, where edge states (solid lines) are formed. The dashed lines represent weak tunneling points. 
With no scatterers inserted the inner edge state is disconnected from the outer one and does not affect 
transport properties. (d) General configuration of the interferometer in the strong field case. The solid 
black circle denotes the transmission coefficient $t$ resulting from a coupling to the inner edge state 
caused by the insertion of an arbitrary phase-coherent scatterer. By unitarity, $t$ is a pure phase. 
Comparing cases (b) and (d) suggests the designation \lq\lq chiral interferometer'' for the latter.}
\label{geometry}
\end{center}
\end{figure}

Resistance measurements have long been used as a spectroscopy of mesoscopic systems, as have other 
spectroscopies such as optical absorption. For example, a measurement of the tunneling current through 
a quantum dot as a function of temperature, voltage, and magnetic field, yields information about the 
electronic many-body states present there. Unfortunately, important information is lost in conventional 
tunneling spectroscopy because only the amplitude $|t|$ of the complex-valued transmission
coefficient $t=|t|e^{i \phi}$ is measured. In a recent series of beautiful experiments, Yacoby {\it et al.} 
\cite{Yacoby etal}, Buks {\it et al.} \cite{Buks etal}, and Schuster {\it et al.} \cite{Schuster etal} 
have succeeded in measuring both the phase and amplitude of the transmission coefficient for tunneling 
through a quantum dot. The phase was measured by inserting a quantum dot into one arm of a mesoscopic 
interferometer ring and observing the shift in the Aharonov-Bohm (AB) magnetoconductance oscillations, 
thereby converting a phase measurement to a multi-probe conductance measurement. The experiments done 
in weak magnetic field used a ring-shaped semiconductor interferometer as shown schematically in 
Fig.~\ref{geometry}a. AB oscillations in the conductance occur as the flux $\Phi$ 
enclosed by the ring 
varies. In Fig.~\ref{geometry}b a phase-coherent scatterer with transmission coefficient $t = |t| 
e^{i \phi}$ is inserted into one arm of the interferometer, resulting in a shift of the phase of the
magnetoconductance oscillations. 

The properties of a ring interferometer in a strong magnetic field are strikingly different than that 
in weak field because of the formation of edge states. Under conditions in which the quantum Hall effect 
is observed, namely, when the Fermi energy in the bulk of the sample is in a mobility gap, the extended 
states responsible for transport lie at the device boundaries \cite{Halperin}. A bare interferometer in 
the quantum Hall regime is shown schematically in Fig.~\ref{geometry}c. The source and drain contacts, 
denoted by the hatched regions, are assumed to be completely phase decoherent. Even in the absence of 
inserted scatterers, the chirality of the edge states dramatically changes the nature of the underlying 
AB interference: First, if there is no coherent transport between the left and right outer edge states, 
there will be no magnetoconductance oscillations at all, because the electrons will travel from source 
to drain without circling flux \cite{absence of AB}. Therefore, weak phase-coherent tunneling points are 
introduced in Fig.~\ref{geometry}c (denoted by dashed lines) to make a viable interferometer, although 
in a real system the coherence length in the contacts might be large enough to observe oscillations. 
Second, in a chiral system the AB oscillations are caused by interference between the direct path from 
source to drain along one edge of the ring and paths containing any number of windings around the ring 
having a given chirality. Whereas in the weak-field case the AB effect leads to both constructive and 
destructive interference (poles and zeros in the probability to propagate around the ring), the AB effect 
in a chiral system therefore leads to constructive interference (poles) only \cite{Geller and Loss}.

We are now in a position to understand the effect of inserting a mesoscopic phase-coherent scatterer, 
such as a quantum point contact or a quantum dot, into one arm of the strong-field interferometer. 
Elastic scattering between the inner and outer edge states is now possible, coupling them together in 
a phase-coherent fashion. Because the coupling to the inner edge state occurs in one arm only, electrons 
scattered to the inner edge state must eventually return to the outer edge state of that same arm. 
Therefore, the effect of any inserted scatterers is to introduce an equivalent scatterer with 
transmission coefficient $t$, shown as a black circle in Fig.~\ref{geometry}d. Usually, $t$ results from 
the transmission through an inserted mesoscopic structure in parallel with the inner edge state of the 
ring. Comparing the equivalent circuits (b) and (d) in Fig.~\ref{geometry}, we see that they are 
distinguished by the chiral nature of the latter. I shall therefore refer to the strong-field ring as a 
{\it chiral interferometer}. An immediate consequence of the chirality is that current conservation 
requires $t$ in case (d) to be a pure phase $e^{i \phi}$. 

The purpose of this paper is to present a brief summary of the rich physics of the chiral interferometer. 
The model I shall adopt here for the interferometer is as follows: Two mesoscopic filling factor 
$g = 1/q$ (with $q$ an odd integer) edge states are coupled to source and drain contacts. Weak 
phase-coherent tunneling points with reflection coefficient $\Gamma_{\rm c}$ (with $|\Gamma_{\rm c}| \ll 1$) 
couple the left and right edge states near the contacts to mimic the residual coherence necessary for 
strong-field interferometry, as discussed above. Because these couplings are assumed to occur in the 
contacts, the coefficients $\Gamma_c$ are assumed not to be renormalized by electron-electron interactions. 
The edges of the two-dimensional electron gas are assumed to be sharply confined, and the interaction 
short-ranged, so that the low lying collective excitations consist of a single branch of 
edge-magnetoplasmons with linear dispersion $\omega = vk$. Then the conductance at zero temperature is 
simply $ G = g \big( 1 - 2 |\Gamma_{\rm c}|^2 [1 + \cos(\theta_{\rm out} +\phi)]\big) {e^2 \over h},$ 
where $\theta_{\rm out}$ is the field-dependent phase accumulated by an electron after traversing the 
outer edge state. I have chosen this model for the bare interferometer because it is the simplest one 
that allows for a measurement of the phase $\phi$; more sophisticated models, including ones where 
$\Gamma_{\rm c}$ is renormalized by interactions, have been studied in a different context elsewhere 
\cite{Geller and Loss,Chamon etal}.  

The dynamics of edge states in the fractional quantum Hall effect regime is governed by Wen's chiral
Luttinger liquid (CLL) theory \cite{Wen review}
\begin{equation}
S_\pm = {1 \over 4 \pi g} \int_0^L \! \! dx \int_0^\beta \! \! d\tau \ \partial_x \phi_\pm \big( \pm i 
\partial_\tau \phi_\pm + v \partial_x \phi_\pm \big),
\label{action}
\end{equation}
where $\rho_{\pm} = \pm \partial_x \phi_{\pm} / 2\pi$ is the charge density fluctuation for right (+) or 
left (--) moving electrons. Canonical quantization in momentum space is achieved by decomposing the chiral 
scalar field $\phi_{\pm}$ into a nonzero-mode contribution $\phi_\pm^{\rm p}$ satisfying periodic 
boundary conditions, and a zero-mode part $\phi_{\pm}^0$. Imposing periodic boundary conditions on the 
bosonized electron field $\psi_{\pm}(x) \equiv (2\pi a)^{-{1\over 2}} e^{iq\phi_\pm(x)} e^{\pm iq\pi x/L}$
($a$ is a microscopic cutoff length)
leads to the requirement that the charge $N_{\pm} \equiv \int_0^L dx \ \rho_{\pm}$ be an integer multiple 
of $g$.

\end{multicols}

The study of mesoscopic effects in the CLL requires a careful treatment of the zero-mode 
dynamics. I shall make extensive use here of the retarded electron propagator $G_\pm(x,t) \equiv -i 
\Theta (t) \langle \lbrace \psi_\pm (x,t) , \psi_\pm^\dagger(0) \rbrace \rangle $ for the finite-size CLL. 
In the presence of an AB flux $\Phi = \varphi \ \! \Phi_0$ (with $\Phi_0 \equiv hc/e$) and additional 
charging energy $U$, the grand-canonical zero-mode Hamiltonian corresponding to (\ref{action}) is
$ H^0_\pm = {1 \over 2} q \Delta \epsilon \big(N_\pm \pm g \varphi)^2 + {1 \over 2} U N_\pm^2 - \mu N_\pm, $ 
where $\Delta \epsilon \equiv 2 \pi v /L$. I then obtain $ \phi_{\pm}^0(x,t) = \pm 2 \pi N_\pm (x \mp vt)/L 
- g \ \! \chi_\pm + g (\mu \mp \varphi \Delta \epsilon)t - g U N_\pm t, $ where $[\chi_{\pm},N_{\pm}]=i,$ 
and (at $T=0$)
\begin{equation}
G_\pm(x,t) = \pm \big({\textstyle {i \over L}}\big)^q \ \! (\pi a)^{q-1} \ \! \Theta(t)
\ \! e^{\pm i q \pi (x \mp vt)/L} \ \! e^{i(\mu \mp \varphi \Delta \epsilon)t}
\big\langle  e^{\pm 2 \pi i q N (x \mp vt)/L} \ \! e^{-iU N t} \big\rangle
\ {\rm Im} \ \! \bigg({ e^{-i U t/2} \over \sin^q \pi(x \mp vt \pm ia)/L } \bigg).
\label{zero temperature G}
\end{equation}

\begin{multicols}{2}
\noindent 
The Fourier transform $G_{\pm}(x,\omega)$ is particularly interesting: For the case $U=0$, it is simply 
related to the Green's function for noninteracting ($q=1$) chiral electrons \cite{mesoscopic CLL}, 
\begin{equation}
G_{\pm}(x,\omega) = G_{\pm}^{q=1}(x,\omega) \times {\epsilon_{\rm F}^{1-q} \over (q-1)! } 
\prod_{j=1}^{q-1} \big(\omega - \omega_j \big) . 
\label{transform}
\end{equation}
Here $\omega_j \equiv [j + {\rm frac}({\mu \over \Delta \epsilon} \mp \varphi)] \ \! \Delta \epsilon$,
where ${\rm frac}(x)$ is the difference between $x$ and its closest integer, and $\epsilon_{\rm F} \equiv
v/a$ is an effective Fermi energy.
Whereas in the $q=1$ case the propagator has poles at each of the $\omega_j$, in the interacting case the first $q-1$ 
poles (above $\mu$) are {\it removed}. 
This effect, which can be regarded as a remnant of the Coulomb blockade for particles with short-range 
interaction, is a consequence of the factor $q$ in the first term of the zero-mode Hamiltonian $H_{\pm}^0$. 
Unlike an ordinary Coulomb blockade, however, the energy gap here, equal to $(q-1)\Delta\epsilon$, is 
exactly quantized. At higher frequencies or in the large $L$ limit where $\omega \gg \Delta \epsilon,$ the 
additional factor becomes $\omega^{q-1} / (q-1)! \ \! \epsilon_{\rm F}^{q-1}$. Upon turning on $U$ a 
conventional Coulomb blockade develops, with a gap given by $U+(q-1)\Delta\epsilon$.

\begin{figure}
\begin{center}
\leavevmode
\epsfbox{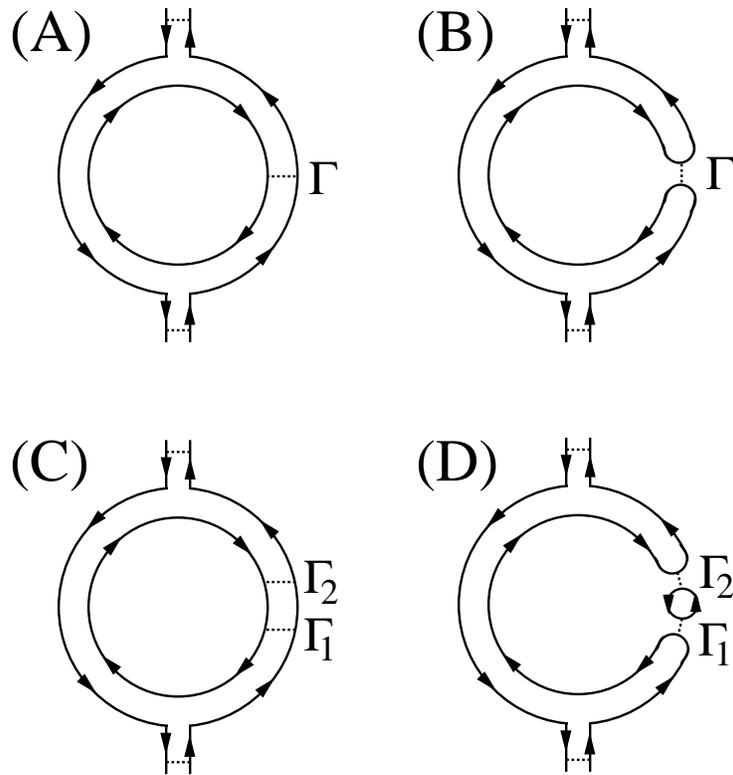}
\caption{Four configurations of the chiral interferometer: (A) One weak tunneling point connecting the 
inner and outer edge states. (B) One strong tunneling point. (C) Two weak tunneling points. (D) A quantum 
dot weakly connected to the incident edge states.}
\label{cases}
\end{center}
\end{figure}

The transmission coefficient for the equivalent scatterer in Fig.~\ref{geometry}d can be 
shown to be given by the {\it ratio} of retarded propagators $t(\epsilon) \equiv G(x_{\rm f},x_{\rm i},\epsilon)/ 
G_{\rm bare}(x_{\rm f},x_{\rm i},\epsilon),$ with $ G_{\rm bare}$ referring to the bare interferometer, which is 
the appropriate generalization of the Fisher-Lee result \cite{Fisher and Lee} to this interacting system.
The proof involves deriving an expression for the source-drain conductance of the interferometer with an
arbitrary inserted scatterer, and extracting the phase shift caused by the latter. 
For the purpose of calculating $t$ we may neglect finite-size effects in the leads and assume
$ G_{\rm bare}(x_{\rm f},x_{\rm i},\epsilon) = G_{\rm bare}(d,\epsilon),$ where $d$ is the size of the inserted
scatterer.   I turn now to a summary of transmission coefficients for the configurations 
shown in Fig.~\ref{cases}; details of the calculations shall be given elsewhere.

(A) {\it Single weak tunneling point}.---I begin with the simple case of one weak tunneling point at $x=x_0$ 
connecting the inner and outer edge states as shown schematically in Fig.~\ref{cases}A. In the fractional 
regime, quasiparticle tunneling, which is allowed in this configuration, diverges at low temperature, driving
the system to the configuration shown in Fig.~\ref{cases}B \cite{Kane and Fisher}. In the integer regime 
$S = S_0 + \delta S$, where 
$S_0 = S_{\rm in} + S_{\rm out}$ is the sum of actions of the form (\ref{action}) for the inner and outer 
edge states, respectively, and $\delta S = \int_0^\beta d\tau \ \! [v \Gamma \psi_{\rm out}(x_0,\tau) 
{\bar \psi}_{\rm in}(x_0,\tau) + {\rm c.c.}]$ is the weak coupling between them. To leading nontrivial order 
perturbation theory yields $t=1+ v^2 |\Gamma |^2 G_{\rm in}(0,\epsilon) G_{\rm out}(a,\epsilon)$,
where $d$ has been taken to be of the order of $a$. The Green's function $G_{\rm in}(x,\omega)$ diverges 
at resonances associated with
the inner edge state, invalidating low-order perturbation theory. However, it is possible to sum the
perturbation expansion to all orders, resulting in 
\begin{equation}
t = {1 + v^2 |\Gamma |^2 G_{\rm in}(0,\epsilon) \big[G_{\rm out}(a,\epsilon) - G_{\rm out}(0,\epsilon)\big] 
\over 1 - v^2 |\Gamma |^2 G_{\rm in}(0,\epsilon) G_{\rm out}(0,\epsilon)}.
\end{equation}
Note that in the CLL it is necessary to distinguish between $G_\pm(a,\epsilon)$ and $G_\pm(0,\epsilon)$, 
because $G_{\pm}(x,\omega)$ is proportional to the unit step function $\Theta( \pm x)$.
At zero-temperature (and $U=0$) a simple expression for the phase shift in this configuration is possible, 
namely $\tan \phi = -{\textstyle {1\over 2}} |\Gamma|^2 {\rm cot} (\theta_{\rm in}/2)
\ \! [1- {1 \over 16} |\Gamma |^4  \ {\rm cot}^2(\theta_{\rm in}/2)]^{-1}  ,$
where $\theta_{\rm in}$ is the phase accumulated by an electron after circling the inner edge 
state \cite{phase footnote}.

\begin{figure}
\begin{center}
\leavevmode
\epsfbox{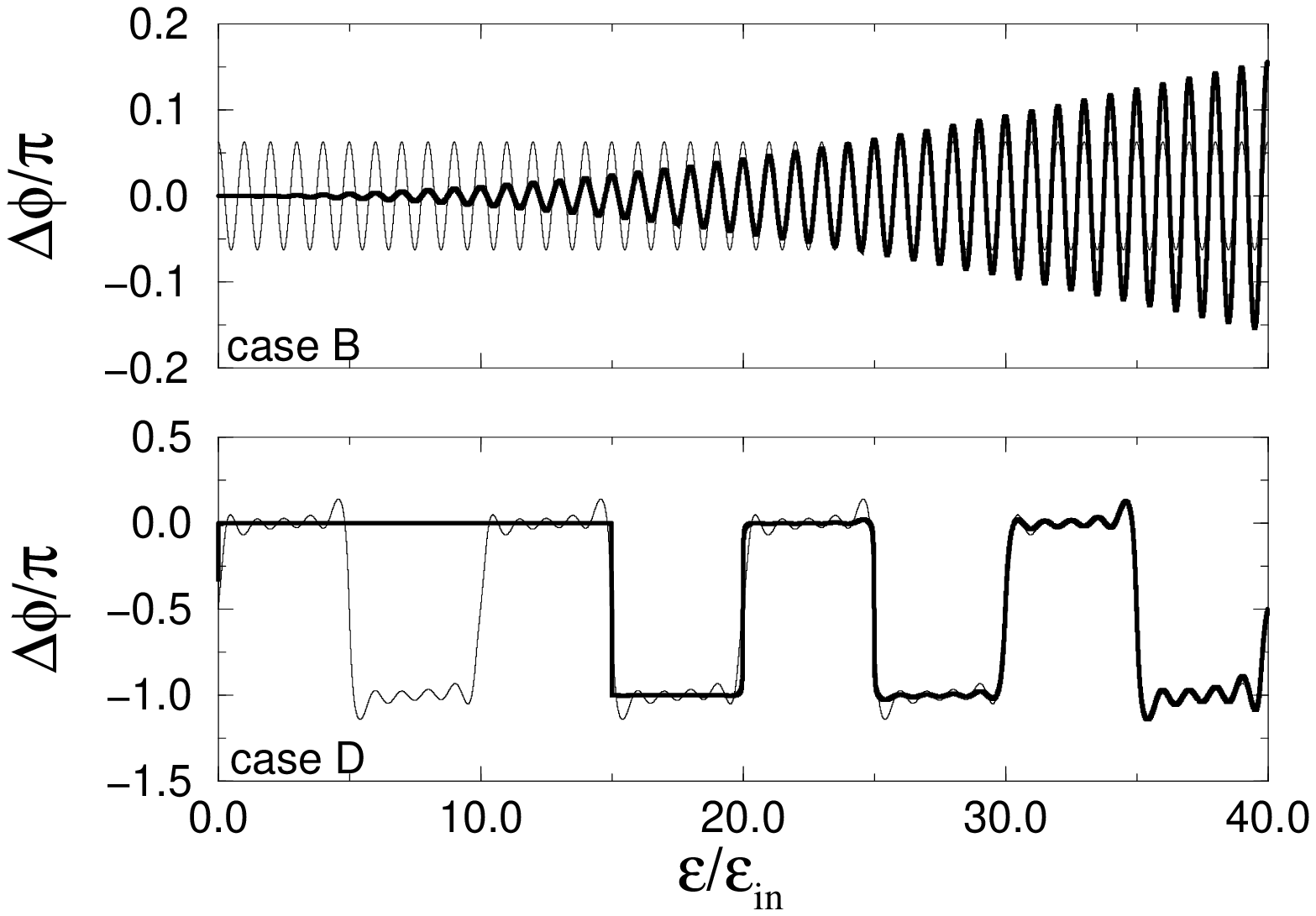}
\caption{Phase of the transmission coefficient as a function of energy for configurations B and D. Here
$\Delta \phi \equiv \phi - 2 \pi \epsilon/\epsilon_{\rm in}$, with $\epsilon_{\rm in} \equiv
2 \pi v / L_{\rm in}$. The thin curves show the case $q=1$ and the thick ones $q=3$. The phase in configuration
D is similar to that in B except for abrupt shifts caused by the quantum dot resonances; in the $q=3$
case the lowest resonances are blocked by interactions [see discussion following Eqn.~(\ref{transform})].}
\label{transmission}
\end{center}
\end{figure}

(B) {\it Single strong tunneling point}.---Next I consider the strong coupling limit of a single 
quantum-point-contact, as shown in Fig.~\ref{cases}B. In this case there is no quasiparticle tunneling. The 
interferometer is described by a single CLL, $S_0 = S_{+}$, taken to be right-moving, and $\delta S = \int_0^\beta 
d\tau \ \! [v \Gamma \psi_{+}(x_1,\tau) {\bar \psi}_{+}(x_2,\tau) + {\rm c.c.}]$. Perturbation theory yields
\begin{equation}
t = { G_{+}(L_{\rm in},\epsilon) - v \Gamma \ \! G_{+}(a,\epsilon)^2 - v \Gamma^* \ \! G_{+}(L_{\rm in},\epsilon)^2 
\over G_{+}(d, \epsilon)}, 
\end{equation}
and, at zero temperature (and $U$ = 0), 
\begin{equation}
\tan \phi = \big[1 + 2 \Gamma (\epsilon/\epsilon_{\rm F})^{q-1} {\rm csc} \ \! (\theta_{\rm in}) 
/(q-1)! \big] \tan \theta_{\rm in},
\label{phase B}
\end{equation}
where $L_{\rm in}$ is the length of the inner edge state. [For simplicity I have assumed in Eqn.~(\ref{phase B}) 
that $\Gamma$ is real and that $d$ is again of the order of $a$.] This expression shows that for $\Gamma = 0$, 
$\phi$ varies linearly with $\epsilon/\epsilon_{\rm in}$ ($\epsilon_{\rm in} \equiv 2 \pi v / L_{\rm in}$) with 
slope $2 \pi$; for finite $\Gamma$ the phase oscillates about this linear variation as shown in 
Fig.~\ref{transmission}.

\end{multicols}

(C) {\it Two weak tunneling points}.---This configuration is similar to that in case A, and to leading order
\begin{equation}
t = 1 + v^2 \big(\sum_i |\Gamma_i|^2 \big) \ \! G_{\rm in}(0,\epsilon) G_{\rm out}(a,\epsilon) + v^2 \Gamma_1 
\Gamma_2^* \ \! {G_{\rm in}(L_{\rm in} \! - d,\epsilon) \ \! G_{\rm out}(a,\epsilon)^2 \over G_{\rm out}
(d,\epsilon)} + v^2 \Gamma_1^* \Gamma_2 \ \! G_{\rm out}(d,\epsilon) \ \! G_{\rm in}(d,\epsilon),
\end{equation}
where $d$ is now the distance between the two quantum point contacts. 

(D) {\it Quantum dot}.---Finally I consider the case of tunneling through a quantum dot weakly coupled to
the interferometer edge states, as shown in Fig.~\ref{cases}D. In this configuration quasiparticle tunneling 
is not allowed, but Coulomb blockade effects are important in the quantum dot. The interferometer is described 
by $S_0 = S_{+} + S_{\rm D}$, where $S_{\rm D}$ is the CLL action for the edge state in the quantum dot 
that includes an additional charging energy $U$, and the weak coupling of the quantum dot to the leads is 
described by $\delta S = \sum_i \int_0^\beta d\tau \ \! [v \Gamma_i \psi_{+}(x_i,\tau) {\bar \psi}_{\rm D}
(x_i,\tau) + {\rm c.c.}]$, with $i=1,2.$ To leading nontrivial order (suppressing the $\epsilon$ dependence
of the Green's functions),
\begin{equation}
t = { G_{+}(L_{\rm in}) \over G_{+}(d) } + v^2 \big(\sum_i |\Gamma_i|^2 \big) \ \! {G_{\rm +}(a) 
\ \! G_{+}(L_{\rm in}) \ \! G_{\rm D}(0) \over  G_{+}(d)} + v^2 \Gamma_1 \Gamma_2^* \ \! {G_{+}(a)^2 
\ \! G_{\rm D}(L_{\rm D}/2) \over  G_{+}(d)} + v^2 \Gamma_1^* \Gamma_2 \ \! {G_{+}(L_{\rm in})^2  
\ \! G_{\rm D}(L_{\rm D}/2) \over G_{+}(d)}, 
\label{case D}
\end{equation}
where $L_{\rm D}$ is the circumference of the quantum dot edge state.
The first term in (\ref{case D}) describes transmission via the inner edge state; the order $|\Gamma_i|^2$
contributions describe the same, apart from an additional tunneling event on and back off the quantum dot at point
$x_i$. The term proportional to $\Gamma_1 \Gamma_2^*$ describes a direct tunneling through the dot, and
the order $\Gamma_1^* \Gamma_2$ term describes transmission via the inner edge state, then backwards
through the quantum dot, and finally around the inner edge state again. The propagator $G_{\rm D}(x,\omega)$
diverges at the quantum dot resonances, invalidating (\ref{case D}), and it is again necessary to sum the
perturbation expansion to all orders; the result (for equal $\Gamma_i$) is
\begin{equation}
t = { G_{+}(L_{\rm in}) + v^2 |\Gamma |^2 \big[ G_{+}(a)^2 \ \! G_{\rm D}(L_{\rm D}/2) 
+ 2 \ \! \Delta \ \! G_{+}(L_{\rm in}) \ \! G_{\rm D}(0) \big] 
+ v^4 |\Gamma |^4 \ \! \Delta^2 \ \! G_{+}(L_{\rm in}) \ \! \big[ G_{\rm D}(0)^2 - G_{\rm D}(L_{\rm D}/2)^2   
\big]  \over G_{+}(d) \big\lbrace 1 - v^2 |\Gamma |^2 \ \! \big[ 2 \ \! G_{+}(0) \ \! G_{\rm D}(0) 
+ G_{+}(L_{\rm in}) \ \! G_{\rm D}(L_{\rm D}/2)] + v^4 |\Gamma |^4 \ \! G_{+}(0)^2 \big[ G_{\rm D}(0)^2
- G_{\rm D}(L_{\rm D}/2)^2 \big] \big\rbrace },
\label{resummed case D}
\end{equation}

\begin{multicols}{2}
\noindent where $\Delta \equiv G_{+}(a,\epsilon)- G_{+}(0,\epsilon).$ The energy-dependent phase for
typical quantum dot parameters is shown in Fig.~\ref{transmission}.

The non-Fermi-liquid nature of the transmission coefficient $t(\epsilon)$ in each configuration manifests 
itself as follows: At a fixed energy $\epsilon$, the phase shift $\phi$ as a function of magnetic field is 
the same as in a Fermi liquid ($q=1$), but the effective coupling constants depend on $\epsilon$. However, 
the energy dependence of $t(\epsilon)$ at fixed field (see Fig.~\ref{transmission}), which can be probed 
by varying the temperature or bias voltage, is dramatically different than in the Fermi liquid case.

It is a pleasure to thank
Hiroshi Akera,
Eyal Buks,
Zachary Ha,
Jung Hoon Han,
Jari Kinaret,
Paul Lammert,
Daniel Loss,
Charles Marcus,
Andy Sachrajda, 
Amir Yacoby,
and Ulrich Z\"ulicke
for useful discussions.

\end{multicols}

\end{document}